\documentclass[11pt]{article}

\usepackage[preprint]{acl}

\usepackage{times}
\usepackage{latexsym}

\usepackage[T1]{fontenc}

\usepackage[utf8]{inputenc}

\usepackage{microtype}

\usepackage{inconsolata}

\usepackage{graphicx}
\usepackage{multirow}
\usepackage{amsmath}
\usepackage{booktabs}       
\usepackage{bm}
\usepackage{graphicx}
\usepackage{amssymb}
\usepackage{soul, xcolor}
\usepackage{algorithm}
\usepackage{algpseudocode}
\usepackage{adjustbox}
\usepackage{siunitx} 

\title{Decoding Ambiguous Emotions with Test-Time Scaling in Audio-Language Models}

\author{
 \textbf{Hong Jia\textsuperscript{1,2}\thanks{Equal contribution.}},
 \textbf{Weibin Li\textsuperscript{3}\footnotemark[1]},
 \textbf{Jingyao Wu\textsuperscript{4}\footnotemark[1]},
 \textbf{Xiaofeng Yu\textsuperscript{2}},
 \\
 \textbf{Yan Gao\textsuperscript{5}},
 \textbf{Jintao Cheng\textsuperscript{3}},
 \textbf{Xiaoyu Tang\textsuperscript{3}},
 \textbf{Feng Xia\textsuperscript{6}},
 \textbf{Ting Dang\textsuperscript{1}},
\\
 \textsuperscript{1}University of Melbourne,
 \textsuperscript{2}University of Auckland,
 \textsuperscript{3}South China Normal University
\\
 \textsuperscript{4}MIT,
 \textsuperscript{5}University of Cambridge
 \textsuperscript{6}RMIT
\\
}

\begin{document}
\maketitle
\begin{abstract}

Emotion recognition from human speech is a critical enabler for socially aware conversational AI. However, while most prior work frames emotion recognition as a categorical classification problem, real-world affective states are often ambiguous, overlapping, and context-dependent, posing significant challenges for both annotation and automatic modeling. Recent large-scale audio language models (ALMs) offer new opportunities for nuanced affective reasoning without explicit emotion supervision, but their capacity to handle ambiguous emotions remains underexplored. At the same time, advances in inference-time techniques such as test-time scaling (TTS) have shown promise for improving generalization and adaptability in hard NLP tasks, but their relevance to affective computing is still largely unknown.
In this work, we introduce the first benchmark for ambiguous emotion recognition in speech with ALMs under test-time scaling. Our evaluation systematically compares eight state-of-the-art ALMs and five TTS strategies across three prominent speech emotion datasets. We further provide an in-depth analysis of the interaction between model capacity, TTS, and affective ambiguity, offering new insights into the computational and representational challenges of ambiguous emotion understanding. Our benchmark establishes a foundation for developing more robust, context-aware, and emotionally intelligent speech-based AI systems, and highlights key future directions for bridging the gap between model assumptions and the complexity of real-world human emotion.

\end{abstract}

\section{Introduction}
Understanding emotion in human conversation is vital for effective communication and social well-being, and it is increasingly important for real-world applications such as mental health support, education, and crisis response~\cite{Elsayed2022, Ramakrishnan2013}. Automated emotion recognition is now deployed in domains including virtual counseling, digital education, and behavioral health assessment~\cite{abd2019overview,vistorte2024integrating,flynn2020assessing}, where systems must accurately and empathetically identify complex emotions to ensure positive outcomes for individuals and communities. However, the majority of current emotion recognition approaches rely on simplifying assumptions, typically categorizing emotions into discrete classes like happiness, sadness, or anger~\cite{niu2024textemotionunveilingemotion, feng2024foundation}. In practice, emotions are often ambiguous, overlapping, or subtle, combining states such as excitement with nervousness or low-intensity blends like sadness and frustration. These nuanced affective expressions are challenging for both annotation and modeling, even for expert humans~\cite{sethu2019ambiguous,wu2024emotion}. As a result, existing models often produce unreliable or misleading outputs, which is of particular concern in applications that serve underrepresented or vulnerable groups.

Recent advances in large language models (LLMs), and in particular audio language models (ALMs) such as Qwen2-Audio~\cite{chu2024qwen2} and Gemini~\cite{team2024gemini}, have introduced new opportunities for nuanced emotion recognition. Thanks to large-scale, multimodal pretraining, ALMs are capable of generating context-rich, human-like descriptions of affective states and show potential to move beyond static category-based approaches. However, these models are typically not trained explicitly for emotion recognition. This raises an important research question: \emph{To what extent can general-purpose ALMs recognize and reason about ambiguous emotions in human speech without explicit emotion supervision?}

A major obstacle to improving emotional intelligence of ALMs in such settings is the scarcity and subjectivity of high-quality ambiguous emotion data for supervised learning~\cite{busso2008iemocap, sethu2019ambiguous, dudzik2024indeterminacy, Mower2009, dang2017investigation, wu24_interspeech}. As an alternative, test-time scaling (TTS)~\cite{snell2024scaling, muennighoff2025s1}, has emerged to scale model performance during the inference phase, without requiring extensive supervised fine-tuning~\cite{snell2024scaling, muennighoff2025s1}. This adaptability makes TTS especially well-suited for tasks like ambiguous emotion recognition, where the complexity and nuance of emotional cues can vary significantly. TTS can potentially enable models to better navigate emotional ambiguity by leveraging their pre-existing knowledge and adjusting to perceptual uncertainties during inference time, while also reducing reliance on costly and hard-to-scale data annotation. 
Given that emotional ambiguity can manifest in varying degrees of intensity and complexity, another critical research question is: \emph{To what extent can TTS enhance ambiguous emotion recognition and, more importantly, dynamically adapt to varying levels of emotional ambiguity?} While much of the existing research has focused on TTS in cognitive intelligence for NLP tasks~\cite{bi2024forest, liu2025can, zhang2025and}, there remains a significant gap in its exploration for ambiguous emotion recognition.

To fill this gap, we present \emph{the first TTS benchmark for ambiguous emotion recognition using audio-language models}. Our contributions are as follows: 
\begin{itemize}
\item We introduce a comprehensive benchmark assessing ambiguous emotion recognition across eight state-of-the-art ALMs and three leading speech emotion datasets. 
\item We systematically evaluate five TTS strategies for ALMs, enabling context-aware, flexible inference in ambiguous emotional scenarios. 
\item We offer in-depth analysis of how TTS effectiveness interacts with varying levels of emotional ambiguity and model scale, providing new insights for the development of emotionally intelligent ALMs. 
\end{itemize}

Our benchmark establishes a foundation for evaluating and advancing emotion recognition systems capable of robustly handling real-world emotional ambiguity, supporting safer and more empathetic AI for high-stakes social applications.
\section{Related Work}
\paragraph{Speech Emotion Recognition using LLMs.}
Speech-based emotion recognition (SER) has long been a central topic in affective computing. Traditional methods have primarily focused on classifying discrete emotion categories or predicting continuous arousal–valence values from speech signals~\cite{yang2018predicting}. The emergence of LLMs and ALMs have opened new avenues for advancing emotion recognition~\cite{feng2024foundation, hu2024exploring}. 
For instance, Feng et al.~\cite{feng2024foundation} fine-tuned LLMs with speech encoders to enable speech-based emotion recognition. Despite these advances, current efforts still rely predominantly on unambiguous, majority-vote labels and often fall short in capturing the subtleties and nuanced variations present in affective cues. 

\paragraph{Ambiguous Emotion Recognition.}
Recent studies have increasingly focused on ambiguous emotion recognition, employing various deep learning approaches for both categorical and dimensional emotion recognition~\cite{wu24_interspeech, dang2017investigation, wu2024handling}, such as Long Short-Term Memory networks~\cite{han2017hard} and Sequential Monte Carlo~\cite{wu2022novel}. With the rise of LLMs, a recent study~\cite{hong2024aerllm} marks an initial step toward ambiguous emotion recognition using LLMs. Their results demonstrate promising performance in cases of low emotional ambiguity but also highlight significant challenges in high-ambiguity scenarios. However, this work is limited to textual inputs and does not explore speech-based models. Another study investigated the use of ALMs with token-level representations and found that token-level representations in ALMs can interpret ambiguous emotions~\cite{halim2025token}. However, this research was limited to a single ALM and lacks a comprehensive evaluation.



\paragraph{Test-Time Scaling.}
Test-time strategies (TTS) enhance model inference without extra training and can be broadly divided into Chain-of-Thought (CoT) prompting~\cite{wei2022chain}, which guides models through intermediate reasoning, and search-with-verifiers~\cite{wang2022self}. 
CoT prompting has demonstrated significant promise in improving the reasoning abilities of large language models (LLMs) by breaking down complex tasks into sequential thought steps~\cite{fu2022complexity}, and recent research has extended its application to multimodal domains, such as audio-visual scene understanding~\cite{shu2023audio}.
The search-with-verifiers approach typically consists of two main stages: generating multiple candidate outputs instead of a single prediction, and then selecting the optimal output using a verifier or reward model. These verifiers can range from simple heuristics to sophisticated, learned reward models. Often, the reward models themselves are fine-tuned LLMs that evaluate, score, and rank the outputs to identify the best response~\cite{adler2024nemotron}. Despite their effectiveness, existing TTC methods have been primarily validated on NLP, coding, and math tasks~\cite{liu2025rethinking}. Their application to audio tasks, particularly ambiguous emotions, remains largely unexplored due to the lack of a clear problem definition and suitable reward models.
\section{Problem Definition}
\begin{figure}[t]
  \centering
  \includegraphics[width=0.5\textwidth]{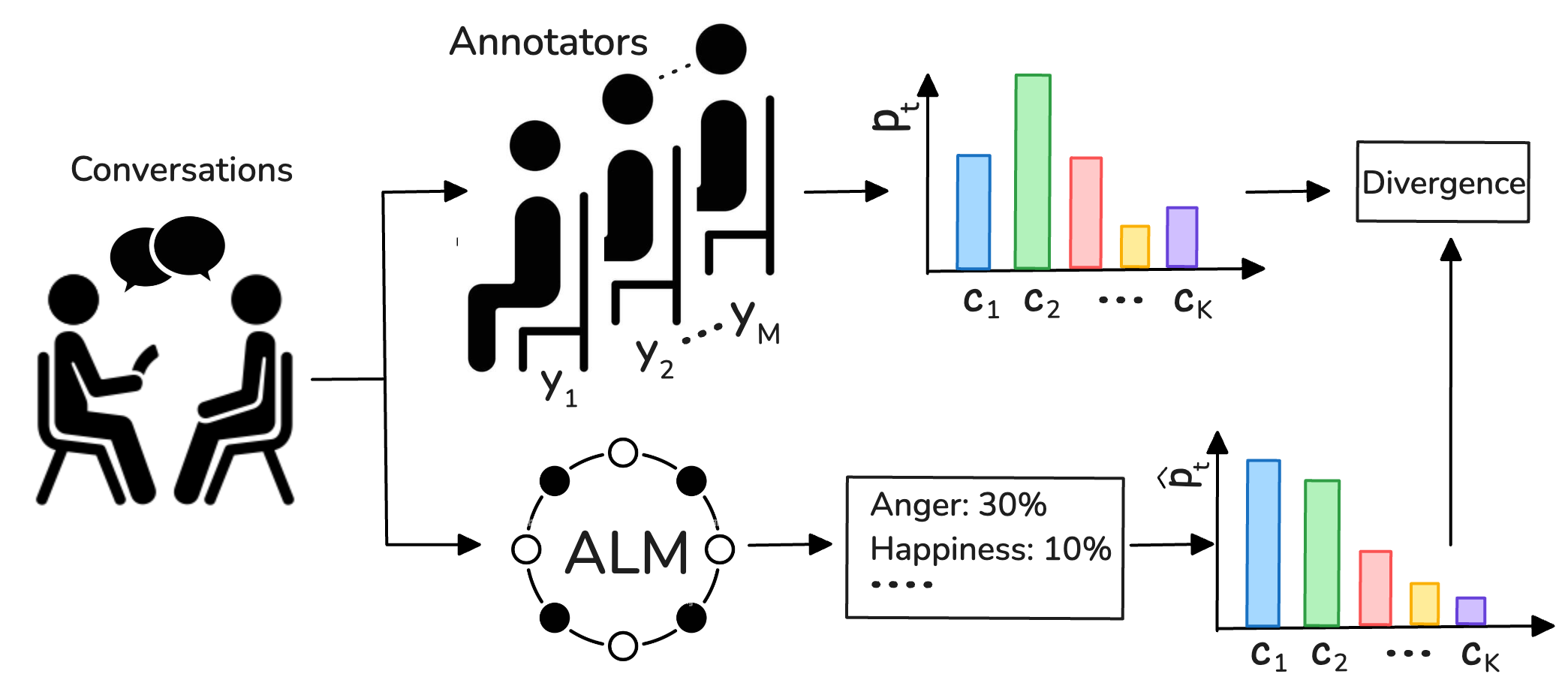}
  \caption{Problem definition of ambiguous emotion recognition using ALMs.}
  \label{fig:prob}
\end{figure}

As illustrated in Figure~\ref{fig:prob}, for a given speech utterance \(\bm{x}_t\), we obtain emotion annotations from \(M\) independent human raters. Each rater assigns a discrete label \(y_t^{(m)} \in \mathcal{C} = \{c_1, \dots, c_K\}\), where \(\mathcal{C}\) represents the predefined set of \(K\) emotion categories. The complete set of annotations for a single utterance is denoted as \(\bm y_t = \{y_t^{(1)}, \dots, y_t^{(M)}\}\). As emotion perception is inherently subjective and ambiguous, annotators frequently disagree; we interpret such disagreement as an indicator of the \emph{emotion ambiguity} for $\bm x_t$.

From these annotations, we derive an empirical emotion distribution \(p_t \in \Delta^{K-1}\) as the emotion labels, where
\begin{equation}
\small
    p_t = \text{SoftLabel}(\bm y_t) = \text{SoftLabel}(y_t^{(1)}, \dots, y_t^{(M)}),
\end{equation}
and \(\Delta^{K-1}\) denotes the \((K\!-\!1)\)-simplex: the set of probability vectors over \(\mathcal{C}\), i.e., \(\sum_{k=1}^K p_{t,k} = 1\), with \(p_{t,k} \geq 0\). The level of disagreement among annotators is measured by the entropy of \(p_t\), which captures the degree of emotional ambiguity. A high entropy value for \(p_t\) indicates greater ambiguity and vice versa. This results in a dataset \(\mathcal{D} = \{(\bm{x}_t, p_t)\}_{t=1}^{T}\), where each instance includes both the speech utterance and its corresponding emotion probability distribution.

The objectives of this work are twofold: first, to evaluate the effectiveness of ALMs $f_\theta: \bm{x} \mapsto \hat{p} \in \Delta^{K-1}$ that predict emotion distributions from speech. 
Second, to enhance emotion recognition and reasoning during inference via TTS, where the model generates $N$ candidate predictions $\{\hat{p}_t^{(1)}, \ldots, \hat{p}_t^{(N)}\}$, and a scoring function~$\phi$ (e.g., reward model) is applied to choose the final output $\hat{p}_t$.
Model predictions are evaluated by quantifying distributional differences from the ground truth, \(\ell(p_t, \hat{p}_t)\), using metrics such as the Jensen–Shannon (JS) divergence. A lower divergence signifies more accurate predictions.

\section{Methods}

\subsection{Overview}
Our benchmarking framework comprises two main components:
(i) Benchmarking existing ALMs for ambiguous emotion recognition (ALM-benchmarking); and
(ii) Investigating TTS for enhancing ambiguous emotion recognition (TTS-benchmarking). The first component investigates the performance of various ALM architectures across multiple datasets, focusing on their ability to recognize ambiguous emotional expressions. The second component centers on assessing the impact of TTS in improving emotion understanding, particularly under varying levels of emotional ambiguity. For the ALM benchmarking, we analyzed eight state-of-the-art ALMs for ambiguous emotion recognition. Detailed prompt designs are provided in the Appendix. In the remainder of this section, we focus on the TTS-benchmarking.

\begin{figure}[t!]
    \centering
    \includegraphics[width=1\linewidth]{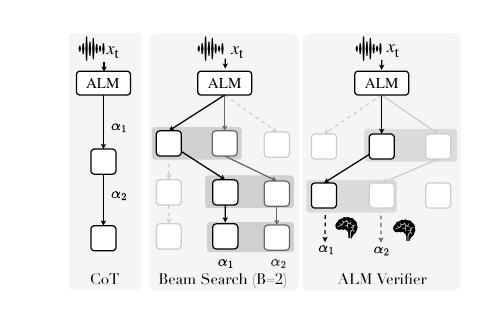}
    \caption{Test-time scaling strategies for ambiguous emotion recognition.} 
    \label{fig:TTS}
\end{figure}

\subsection{TTS methods}
Five different TTS methods have been evaluated: CoT prompting~\cite{wang2022self} and four search against verifiers, as shown in Figure~\ref{fig:TTS}.
\subsubsection{CoT prompting} We incorporate CoT prompting for ambiguous emotion recognition, leveraging prior knowledge of speech emotional cues to guide the LLMs through a step-by-step reasoning process. The specifics of the CoT prompting are detailed in the Appendix.


\subsubsection{Beam search with Best of N} The Best-of-N method generates multiple responses for a given audio input $\bm{x}_t$. This technique often employs beam search to explore several decoding paths. At each decoding step, beam search maintains the top $B$ candidate sequences, enabling the model to consider a range of high-probability outputs and thereby enhancing both diversity and accuracy.

At each decoding step, the algorithm expands all possible next tokens for each current sequence and retains only the top $B$ sequences with the highest cumulative log probabilities. This process, known as beam search, keeps a "beam" of the $B$ most likely candidate sequences at each time step and iteratively extends them until completion. After decoding, the final set of $B$ candidate sequences is typically evaluated by a reward model to assess their quality. However, as reward models specifically designed for ALMs and emotion tasks are still lacking, we propose to use the cumulative log-likelihood $\alpha_b,\, b \in [1, B]$ as the scoring metric for selecting the final output.

Specifically, we consider two scoring strategies: i) \emph{Maximum Log-Likelihood:} select the candidate with the highest log-likelihood as $\hat{p}_t^* = \hat{p}_t^{\left( \arg\max_{b} \alpha_b \right)}$; and ii) \emph{Weighted Combination:} compute a weighted sum of all outputs generated by beam search, where each output $\hat{p}_t^{(b)}$ is weighted by its normalized log-likelihood $\alpha_b$. 

To aggregate the \( B \) predicted empirical emotion distributions \( \hat{p}_t^{(b)} \), we use a Dirichlet Mixture Model (DMM). Each prediction \( \hat{p}_t^{(b)} \) serves as the mean of a Dirichlet distribution with concentration parameters \( \boldsymbol{\beta}_b = \tau\, \hat{p}_t^{(b)} \), where the scaling factor \( \tau \) controls sharpness. We set \( \tau = 1 \) to ensure the resulting distributions are well-formed probability distributions (i.e., their components sum to 1).

To reflect the relative importance of each prediction, we weight the Dirichlet components using the normalized log-likelihoods as above:
\begin{equation}
w_b = \frac{\exp(\alpha_b)}{\sum_{j=1}^B \exp(\alpha_j)}.
\end{equation}
The expected aggregated distribution under the DMM is:
\begin{equation}\label{eq:weighted}
\small
\hat{p}_t^* = \sum_{b=1}^B w_b\, \mathbb{E}_{\mathrm{Dir}(\boldsymbol{\beta}_b)}[\boldsymbol{p}] = \sum_{b=1}^B w_b\, \frac{\boldsymbol{\beta}_b}{\sum_{k} \beta_{b,k}} = \sum_{b=1}^B w_b p_t^{(b)},
\end{equation}
where \( \mathbb{E}_{\mathrm{Dir}(\boldsymbol{\beta}_b)}[\boldsymbol{p}] \) denotes the mean of the Dirichlet distribution.
This is mathematically equivalent to a weighted sum, but the DMM provides a principled way to aggregate predictions while explicitly accounting for both relative importance and uncertainty in the distributions.

The weighted combination allows the model to explore more possible outputs, which increases diversity in what it generates. This can be especially beneficial for processing complex or uncertain speech emotion, where capturing a wide range of possible outputs is advantageous.

\subsubsection{ALM verifier } An alternative method to score the top $B$ beams is to use a more powerful ALM as a verifier. This verifier is selected for its enhanced audio comprehension and emotion recognition, enabling it to assess the responses of the primary model more effectively.

The verifier takes the audio input and each of the $B$ responses, along with a prompt describing the evaluation task and scoring criteria. It then assigns a score to each response. The final output is chosen as the response with the highest score, or alternatively, a weighted combination of all responses can be computed based on the verifier's scores, similar as in Eq~\eqref{eq:weighted}.
\section{Experimental Setups}
\paragraph{Dataset. }
We evaluate our methods on three widely used emotion recognition datasets: IEMOCAP~\cite{busso2008iemocap}, MSP-Podcast~\cite{lotfian2017building}, and CREMA-D~\cite{cao2014crema}. These datasets collectively offer a diverse range of speaking styles, emotional expressions, and annotation protocols, making them suitable for studying ambiguous emotion recognition in speech. IEMOCAP~\cite{busso2008iemocap} is a multimodal dataset containing approximately 12 hours of audiovisual recordings from dyadic sessions between actors. Each utterance is annotated by multiple raters using categorical emotion labels. For our experiments, we focus on utterances that were labeled with four commonly used categories: happy, angry, sad, and neutral, leading to 4,373 utterances for analysis. 

MSP-Podcast~\cite{lotfian2017building} consists of naturalistic speech segments extracted from podcast recordings, offering a more ecologically valid emotional distribution. Eight emotions were selected for evaluations, leading to 12,955 utterances. CREMA-D~\cite{cao2014crema} is a controlled dataset comprising over 7,400  audio-visual clips of 91 actors expressing six basic emotions across different intensities. Each clip is annotated by a crowd of raters.

\paragraph{Implementation details. }
Eight different audio-language models have been evaluated, including open-source models such as Audio-Flamingo 2~\cite{ghosh2025audio}, Qwen2.5-Omni~\cite{xu2025qwen2}, Qwen2-Audio-Instruct~\cite{Qwen2-Audio},  and two versions of Ultravox series from Fixie Ai~\cite{ultravox_v03,ultravox_v04}, as well as closed-source models like Gemini2.5-pro and Gemini2.0-flash ~\cite{team2023gemini} and GPT-4o~\cite{achiam2023gpt}. Details of the model structures can be found in the Appendix. 

Five TTS methods are evaluated: \emph{CoT}, \emph{Best-of-N (BoN)} which selects the top beam from beam search, \emph{Weighted BoN (W-BoN)} as defined in Eq.~\eqref{eq:weighted},
\emph{ALM verifier (ALM-v)} which employs GPT-4o to choose the best response due to its strong performance in general NLP tasks, \emph{Weighted ALM verifier (W-ALM-v)}, which uses the scores of stronger ALMs as weights. We used $B=5$ for \emph{BoN} and \emph{W-BoN}, and $B=3$ for \emph{ALM-v} and \emph{W-ALM-v} respectively, with the beam size optimized within $[2,7]$\footnote{The code will be made available upon acceptance.}. 


\begin{table*}[t]
    \centering
    \label{tab:alms_revised}
    \resizebox{\linewidth}{!}{%
    \begin{tabular}{l *{9}{S[table-format=1.4]}}
        \toprule
        & \multicolumn{3}{c}{\textbf{IEMOCAP}} & \multicolumn{3}{c}{\textbf{MSP-Podcast}} & \multicolumn{3}{c}{\textbf{CREMA-D}} \\
        \cmidrule(lr){2-4} \cmidrule(lr){5-7} \cmidrule(lr){8-10}
        \textbf{Model} & {JS $\downarrow$} & {BC $\uparrow$} & {$R^2$ $\uparrow$} & {JS $\downarrow$} & {BC $\uparrow$} & {$R^2$ $\uparrow$} & {JS $\downarrow$} & {BC $\uparrow$} & {$R^2$ $\uparrow$} \\
        \midrule
        \multicolumn{10}{l}{\textit{Open-Source Models}} \\
        \midrule
        Audio-Flamingo 2 (3B) & 0.4920 & 0.5282 & 0.4151 & 0.6206 & 0.3908 & 0.2008 & 0.4539 & 0.5672 & 0.4756 \\
        Qwen2-Audio (7B)      & 0.3592 & \textbf{0.6477} & 0.3559 & 0.4128 & 0.6123 & 0.3601 & 0.3172 & 0.7051 & 0.3959 \\
        Qwen2.5-Omni (7B)     & 0.3855 & 0.5794 & 0.4664 & 0.4957 & 0.5039 & 0.3689 & 0.4159 & 0.6078 & 0.4875 \\
        Ultravox-v0.3 (8B)    & 0.3694 & 0.5843 & 0.3973 & 0.4499 & 0.5725 & 0.3425 & 0.4438 & 0.5899 & 0.4296 \\
        Ultravox-v0.4 (8B)    & 0.3717 & 0.5831 & 0.3956 & 0.4485 & 0.5758 & 0.3448 & 0.4400 & 0.5927 & 0.4336 \\
        \midrule
        \multicolumn{10}{l}{\textit{Closed-Source Models}} \\
        \midrule
        GPT-4o    & \underline{0.3385} & 0.6217 & 0.4837 & \underline{0.3737} & \underline{0.6584} & 0.3964 & \textbf{0.2935} & \underline{0.7434} & \underline{0.4876} \\
        Gemini 2.0 Flash      & 0.3501 & 0.6163 & \textbf{0.5148} & 0.3980 & 0.6446 & \underline{0.4057} & 0.3439 & 0.7082 & 0.4837 \\
        Gemini 2.5 Pro        & \textbf{0.3378} & \underline{0.6332} & \underline{0.5056} & \textbf{0.3710} & \textbf{0.6716} & \textbf{0.4148} & \underline{0.2983} & \textbf{0.7520} & \textbf{0.5296} \\
        \bottomrule
    \end{tabular}
    }
    \caption{
        Comparison of ALMs for ambiguous emotion recognition on three benchmark datasets. 
        The best and second-best results in each column are highlighted in \textbf{bold} and \underline{underlined}.
    }
    \label{tab:alms}
\end{table*}

\paragraph{Evaluation. }
The key evaluation measures the similarity between the predicted and ground truth emotion distributions. We use Jensen-Shannon Divergence (JS), Bhattacharyya Coefficient (BC), and $R^2$ as primary evaluation metrics. JS divergence quantifies the difference between the predicted and actual distributions (lower is better), BC assesses distributional similarity (higher is better), and $R^2$ evaluates goodness-of-fit (higher is better). 
All metrics range from 0 to 1. Although the model is optimized for ambiguous emotions in terms of distributions, a good prediction should also align the dominant emotion with the majority vote of the annotations. Therefore, we additionally compare the predicted dominant emotion (e.g., one with the highest predicted probability) to the annotators' majority vote, using accuracy and F1-scores.

\section{Results}
\subsection{Ambiguous Emotion Recognition}
Table~\ref{tab:alms} presents the ambiguous emotion recognition performance of eight open-source and closed-source models.
Closed-source models typically demonstrate superior performance compared to open-source models. Gemini 2.5 Pro consistently achieves the best or near-best results across almost all datasets and metrics. Notably, it achieves the best results across all datasets and metrics, with the exception of the BC and $R^2$ on IEMOCAP and the JS on CREMA-D, where it secures the second-best scores. GPT-4o and Gemini 2.0 Flash also perform strongly; GPT-4o sets a new low JS for CREMA-D ($0.2935$), indicating the closest class distribution match on that particular dataset.

Among open-source models, Qwen2-Audio (7B) is the top performer, achieving the lowest JS and highest BC on the IEMOCAP, MSP-Podcast, and CREMA-D datasets. Notably, it attains the highest BC on IEMOCAP (0.6477), surpassing even the closed-source models. Qwen2.5-Omni (7B) achieves the highest $R^2$ on IEMOCAP, MSP-Podcast, and CREMA-D datasets among the open-source models, suggesting superior ability to capture the variance inherent in those datasets. In contrast, both Ultravox-v0.3 (8B) and Ultravox-v0.4 (8B) exhibit similar, but generally lower, performance on all metrics compared to the Qwen2.

While Gemini 2.5 Pro excels, Qwen2-Audio narrows the gap on IEMOCAP and MSP-Podcast, demonstrating that high-performing open-source alternatives are becoming increasingly competitive. It is also notable that model scale, as measured in billions of parameters, does not always translate directly into better performance, as evidenced by the competitive results of the 7B Qwen2 models relative to the larger 8B Ultravox models.

In sum, these results reaffirm that current models can recognize ambiguous emotions to a certain degree, and both closed-source and open-source models demonstrate substantial potential. 

\begin{table}[t!]
    \centering
    \resizebox{\columnwidth}{!}{%
    \begin{tabular}{llll}
        \toprule
        \multicolumn{4}{c}{\textbf{IEMOCAP}} \\
        \cmidrule(lr){1-4}
        \textbf{Method} & \multicolumn{1}{c}{JS $\downarrow$} & \multicolumn{1}{c}{BC $\uparrow$} & \multicolumn{1}{c}{$R^2$ $\uparrow$} \\
        \midrule
        w/o TTS         & \multicolumn{1}{c}{0.3592} & \multicolumn{1}{c}{0.6477} & \multicolumn{1}{c}{0.3559} \\
        \midrule
        CoT             & 0.3578 (+0.4\%) & 0.6409 (-1.0\%)  & 0.3792 (+6.5\%) \\
        BoN             & 0.3450 (+4.0) & 0.6781 (+4.7)  & \textbf{0.3815 (+7.2)} \\   
        W-BoN           & \textbf{0.3364 (+6.3)} & \textbf{0.6915 (+6.8)} & 0.3742 (+5.1) \\
        ALM-v           & 0.3437 (+4.3) & 0.6817 (+5.2)  & \underline{0.3794 (+6.6)} \\
        W-ALM-v         & \underline{0.3433 (+4.4)} & \underline{0.6853 (+5.8)} & 0.3749 (+5.3) \\
        \midrule
        \midrule
        \multicolumn{4}{c}{\textbf{MSP-Podcast}} \\
        \cmidrule(lr){1-4}
        \textbf{Method} & \multicolumn{1}{c}{JS $\downarrow$} & \multicolumn{1}{c}{BC $\uparrow$} & \multicolumn{1}{c}{$R^2$ $\uparrow$} \\
        \midrule
        w/o TTS         & \multicolumn{1}{c}{0.4128} & \multicolumn{1}{c}{0.6123} & \multicolumn{1}{c}{0.3601} \\
        \midrule
        CoT             & 0.4393 (-6.4) & \underline{0.6138 (+0.2)} & 0.3179 (-11.7) \\
        BoN             & 0.4600 (-11.4)& 0.5636 (-8.0) & 0.3410 (-5.3) \\
        W-BoN           & 0.3960 (+4.1) & \textbf{0.6207} (\textbf{+1.4}) & 0.3642 (+1.1) \\
        ALM-v           & \underline{0.2949 (+28.6)} & 0.5945 (-2.9) & \textbf{0.3763 (+4.5)} \\
        W-ALM-v         & \textbf{0.2845} (\textbf{+31.1}) & 0.6110 (-0.2)  & \underline{0.3669 (+1.9)} \\
        \midrule
        \midrule
        \multicolumn{4}{c}{\textbf{CREMA-D}} \\
        \cmidrule(lr){1-4}
        \textbf{Method} & \multicolumn{1}{c}{JS $\downarrow$} & \multicolumn{1}{c}{BC $\uparrow$} & \multicolumn{1}{c}{$R^2$ $\uparrow$} \\
        \midrule
        w/o TTS         & \multicolumn{1}{c}{0.3172} & \multicolumn{1}{c}{0.7051} & \multicolumn{1}{c}{0.3959} \\
        \midrule
        CoT             & 0.3310 (-4.4) & 0.7022 (-0.4) & 0.3549 (-10.3) \\
        BoN             & 0.3323 (-4.8) & 0.6849 (-2.9) & 0.4219 (+6.6) \\
        W-BoN           & \textbf{0.2908 (+8.3)} & \textbf{0.7286 (+3.3)}  & \underline{0.4559 (+15.2)} \\
        ALM-v           & 0.3066 (+3.3) & 0.7139 (+1.2)   & \textbf{0.4645 (+17.3)} \\
        W-ALM-v         & \underline{0.2972 (+6.3)} & \underline{0.7245 (+2.8)} & 0.4500 (+13.7) \\
        \bottomrule
    \end{tabular}%
    }
    \caption{Comparison of TTS methods using Qwen2-Audio on three benchmark datasets. The best and second-best results for each dataset are highlighted in \textbf{bold} and \underline{underlined}. Relative changes are shown in brackets in percentage (\%).}
    \label{tab:TTS}
\end{table}

\begin{figure*}[t]
  \centering
  \includegraphics[width=1\textwidth]{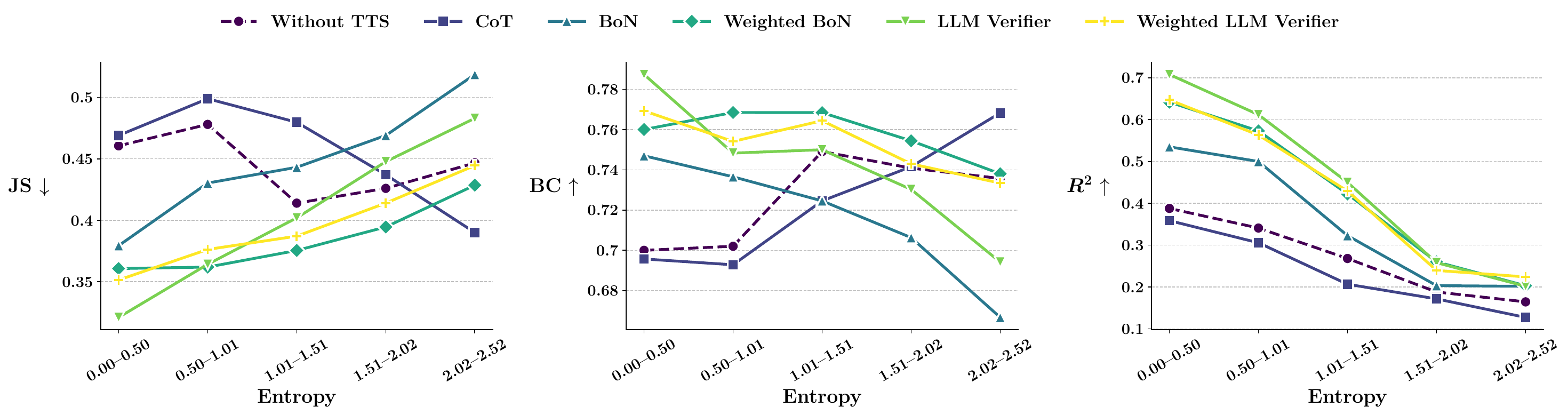}
  \caption{TTS performance with varying levels of ambiguity from low to high, quantified by entropy on CREMA-D.}
  \label{fig:TTS_line_plot}
\end{figure*}

\subsection{Performance with TTS}
Table \ref{tab:TTS} provides a comparative evaluation of different TTS strategies using Qwen2-Audio, which demonstrated the best performance among open-source models. Closed-source models are not considered, as they do not provide the inference flexibility required for beam search in TTS. Overall, TTS strategies significantly impact ambiguous emotion recognition, though their effectiveness varies across datasets. On IEMOCAP, BoN consistently outperforms the baseline across all metrics, whereas CoT shows mixed results, improving JS and $R^2$ but slightly degrading BC. 

Among all methods, Weighted BoN and Weighted ALM verifier stand out with strong and consistent performance across metrics and datasets. Particularly, weighted BoN achieves the highest BC on all three datasets, with relative improvements up to 6.8\%. 
It also yields the second highest on CREMA-D for $R^2$. 
Additionally, it improves JS by 4.1\% and 6.3\% on MSP-Podcast and IEMOCAP.

Weighted ALM verifier achieves the lowest JS divergence on MSP-Podcast, with a 31.1\% relative reduction, and shows consistent improvements across other metrics. On IEMOCAP, it secures the second-best scores. Similarly, on CREMA-D, it ranks second in both metrics. 


Interestingly, unweighted BoN exhibits performance degradation on MSP-Podcast. This likely reflects the dataset's spontaneous and subtle emotional nature, where selecting a single 'best' beam fails to capture the full distribution. In contrast, aggregation strategies (W-BoN and W-ALM-v) substantially reverse this trend. For instance, W-BoN achieves the best overall balance, achieving the lowest JS (0.2908) on CREMA-D and the highest BC (0.6207) on MSP-Podcast. This confirms that weighing multiple plausible outputs based on confidence scores, rather than discarding them, is critical for modeling emotional ambiguity.

These findings indicate that relying on a single prediction may fail to capture the nuanced and multifaceted nature of human emotions; however, incorporating multiple diverse outputs enables the model to better account for this ambiguity. In general, the overall improved performance suggests that \emph{aggregating multiple plausible responses by exploring a broader latent space can significantly improve model performance, as it allows the foundational knowledge learned during pretraining to be fully leveraged during inference}. This approach is particularly valuable in the context of ambiguous emotion recognition with varying uncertainty.

\subsection{TTS with Varying Ambiguity Levels}
To better understand how TTS improves emotion recognition across different levels of ambiguity, we further assess its performance under varying degrees of ambiguity. Ambiguity levels are determined by the entropy of the label distributions, with higher entropy reflecting more ambiguous emotions and greater annotator disagreement.

The ambiguity levels, in terms of entropy, are categorized into five equal groups. We compared the TTS performance across these levels using the CREMA-D dataset in Figure \ref{fig:TTS_line_plot}. This analysis reports the median values of the metrics to reduce the impact of occasional outliers on the overall evaluation of model performance.  

\paragraph{JS divergence.} The base model without TTS performs unexpectedly poorly on low-entropy (less ambiguous) emotions, while it performs better on more ambiguous emotions. This may be attributed to the training scheme of Qwen2-Audio, which focuses on ``general audio understanding'' by incorporating over 30 diverse audio types, including non-speech sounds and music in its training data and tasks. Such data diversity likely leads to richer representation learning associated with emotion. Across most TTS methods, JS divergence generally increases as entropy rises, suggesting that as emotional ambiguity increases, model predictions deviate further from the ground truth distribution, highlighting the difficulty of high-entropy data. 

Notably, despite this upward trend, both weighted BoN and weighted LLM Verifier consistently outperform the base model across all entropy levels. These consistent improvements indicate that leveraging multiple beam paths can enhance inference quality. Interestingly, CoT exhibits a different pattern from other TTS methods. It shows significant advantages as entropy increases, suggesting that step-by-step reasoning in CoT can more effectively interpret ambiguous emotions. As a result, CoT achieves the best performance in high-entropy situations, though it is less effective in low-entropy contexts.

Overall, the results indicate that various TTS methods can improve recognition of ambiguous emotions across all entropy regions, though the degree of improvement varies depending on both the TTS method used and the entropy level.

\paragraph{BC and $\bm R^2$. }For BC, we observed the same conclusions as with JS divergence. Regarding $R^2$, all TTS methods show a decreasing trend, reflecting increased difficulty in accurately estimating the distribution as entropy rises. Nonetheless, weighted BoN, the LLM Verifier, and the weighted LLM Verifier consistently deliver improvements over the base models. Detailed performance of ALMs with varying ambiguity levels can be found in the Appendix.




\subsection{Single-Class Emotion Recognition}
To further analyze how the predicted distributions preserve the dominant emotion, we compared the predicted dominant emotion with the majority vote from the annotations. The predicted dominant emotion is identified as the category with the highest probability in the predicted distribution. 

\paragraph{ALMs. }Table \ref{tab:majority_vote} presents the performance of ALMs without TTS. Closed-source models such as the Gemini series and GPT-4o consistently outperform others across the datasets. Gemini 2.5 Pro achieves the highest accuracy on both the IEMOCAP and CREMA-D datasets, whereas GPT-4o excels on MSP-Podcast. This observation aligns with the overall distribution predictions, suggesting that the emotion distributions predicted by these models can also capture the dominant emotion.

Among the open-source models, Qwen2.5 stands out, whereas Qwen2-Audio performs poorly in identifying the dominant emotion in the IEMOCAP and MSP-Podcast datasets. This indicates that different Qwen versions possess varying capabilities: Qwen2.5 is more adept at detecting the dominant emotion, while Qwen2-Audio better perceives the overall emotion distribution.

Notably, on the CREMA-D dataset, several models achieved reasonable accuracy but exhibited extremely low macro-F1 scores. For example, GPT-4o attained an accuracy of 0.5418 but only an F1-score of 0.1231. Our analysis indicates that this discrepancy is due to the failure in predicting the ``disgust'' emotion, which accounts for 9.1\% of the data. This suggests that all evaluated ALMs currently lack the capability to recognize ``disgust,'' highlighting an area for future improvement. Interestingly, Qwen2-Audio was unique in its partial success at predicting ``disgust,'' resulting in the highest F1-score on CREMA-D.

\paragraph{ALMs with TTS.}Table~\ref{tab:TTS_qwen2_majority_vote} shows that the base model without TTS underperforms consistently in both accuracy and F1 score. Most TTS methods improve over the base model, with the weighted methods of W-BoN and W-LLM verifier delivering the best overall results.
Specifically, W-BoN achieves the highest F1 on IEMOCAP (0.3688) and leads in both accuracy (0.4225) and F1 (0.2373) on MSP-Podcast, while W-LLM verifier performs best in F1 on CREMA-D (0.3630). However, methods such as CoT and BoN occasionally underperform the baseline model, particularly on the MSP-Podcast dataset which features more naturalistic emotional expressions. This suggests that when these reasoning strategies, optimized for exploring emotional ambiguity, are constrained to a single-label classification task, their advantages may not translate effectively and could even introduce interference. In summary, these results demonstrate that while weighted aggregation strategies consistently enhance emotion classification performance across datasets, the effectiveness of different TTS methods remains highly dependent on the specific task and data characteristics.

\begin{table}[t!]
    \centering
    \small
    \resizebox{\columnwidth}{!}{%
    \setlength{\tabcolsep}{1.6pt}
    \begin{tabular}{l S[table-format=1.4] S[table-format=1.4] S[table-format=1.4] S[table-format=1.4] S[table-format=1.4] S[table-format=1.4]}
        \toprule
        & \multicolumn{2}{c}{\textbf{IEMOCAP}} & \multicolumn{2}{c}{\textbf{MSP-Podcast}} & \multicolumn{2}{c}{\textbf{CREMA-D}} \\
        \cmidrule(lr){2-3} \cmidrule(lr){4-5} \cmidrule(lr){6-7}
        \textbf{Model} & {Acc} & {F1} & {Acc} & {F1} & {Acc} & {F1} \\
        \midrule
        \multicolumn{7}{l}{\textit{Open-Source Models}} \\
        \midrule
        Audio-Flamingo\hspace{-0.05em} 2 & 0.3356 & 0.3071 & 0.2509 & 0.1217 & 0.5216 & 0.1715 \\
        Qwen2-Audio      & 0.2913 & 0.2061 & 0.3830 & 0.2142 & 0.3670 & \textbf{0.3289} \\
        Qwen2.5          & 0.4577 & 0.4246 & 0.4080 & 0.2643 & 0.5402 & 0.1341 \\
        Ultravox-v0.3    & 0.3894 & 0.3394 & 0.3902 & 0.2348 & 0.4676 & 0.1224 \\
        Ultravox-v0.4    & 0.3878 & 0.3422 & 0.3969 & 0.2379 & 0.4747 & 0.1238 \\
        \midrule
        \multicolumn{7}{l}{\textit{Closed-Source Models}} \\
        \midrule
        GPT-4o           & 0.4460 & 0.4227 & \textbf{0.6236} & \textbf{0.3657} & \underline{0.5418} & 0.1231 \\
        Gemini 2.0 flash     & \underline{0.4580} & \underline{0.4417} & 0.4337 & 0.2666 & 0.5357 & 0.1247 \\
        Gemini 2.5 pro       & \textbf{0.4697} & \textbf{0.4588} & \underline{0.4372} & \underline{0.2885} & \textbf{0.5927} & \underline{0.2555} \\
        \bottomrule
    \end{tabular}
    }
    \caption{
        Performance comparison of various ALMs on single-class emotion classification. 
    }
    \label{tab:majority_vote}
\end{table}

\begin{table}[t!]
    \centering
    \small
    \resizebox{\columnwidth}{!}{%
    \setlength{\tabcolsep}{3.8pt}
    \begin{tabular}{c S[table-format=1.4] S[table-format=1.4] S[table-format=1.4] S[table-format=1.4] S[table-format=1.4] S[table-format=1.4]}
        \toprule
        & \multicolumn{2}{c}{\textbf{IEMOCAP}} & \multicolumn{2}{c}{\textbf{MSP-Podcast}} & \multicolumn{2}{c}{\textbf{CREMA-D}} \\
        \cmidrule(lr){2-3} \cmidrule(lr){4-5} \cmidrule(lr){6-7}
        \textbf{Method} & \multicolumn{1}{c}{Acc} & \multicolumn{1}{c}{F1} & \multicolumn{1}{c}{Acc} & \multicolumn{1}{c}{F1} & \multicolumn{1}{c}{Acc} & \multicolumn{1}{c}{F1} \\
        \midrule
        w/o TTS                 & 0.2913 & 0.2061 & 0.3830 & 0.2142 & 0.3670 & 0.3289 \\
        \midrule
        CoT                     & 0.3310 & 0.3218 & 0.3357 & 0.1896 & 0.2999 & 0.3091 \\
        BoN                     & \textbf{0.3729} & \underline{0.3637} & 0.3461 & 0.2093 & 0.4391 & 0.3258 \\
        W-BoN                   & \underline{0.3669} & \textbf{0.3688} & \textbf{0.4225} & 0.2373 & \textbf{0.5324} & 0.3319 \\
        ALM-v            & 0.3386 & 0.3186 & \underline{0.4171} & \textbf{0.2533} & 0.5126 & \underline{0.3523} \\
        W-ALM-v          & 0.3431 & 0.3267 & 0.4131 & \underline{0.2522} & \underline{0.5230} & \textbf{0.3630} \\
        \bottomrule
    \end{tabular}
    }
    \caption{
        Performance comparison of various TTS methods for single-class emotion classification using Qwen2-Audio.
    }
    \label{tab:TTS_qwen2_majority_vote}
\end{table}

\section{Conclusion}
In this work, we systematically evaluate ALMs and various TTS methods for ambiguous emotion recognition across multiple benchmark datasets. Our results clearly demonstrate that ALMs are capable of recognizing ambiguous emotions to a certain extent, with closed-source models generally performing better. When incorporating TTS, performance improves significantly, particularly with weighted aggregation strategies such as W-BoN and W-LLM verifier, which substantially outperform other TTS approaches. These findings underscore the importance of considering multiple solutions to address emotion ambiguity. Furthermore, the effectiveness of TTS methods varies depending on the dataset and the degree of ambiguity present.


\section{Limitations}

While our findings indicate that test-time scaling (TTS) enables audio language models to move beyond rigid classification and exhibit greater uncertainty awareness in the presence of subtle or blended emotions, several limitations remain.

First, Our evaluation is currently bounded by the available datasets and annotation schemes, which do not yet reflect the full diversity and complexity of natural emotional expression across cultures, speakers, and contexts. Expanding future benchmarks to include more varied, multilingual data, richer annotation formats such as annotator distributions, confidence scores, or continuous emotion representations will offer even greater insights and realism when the data is available.

Additionally, our work used general-purpose TTS methods. There is exciting potential to develop emotion-focused TTS strategies that dynamically adjust inference for ambiguous cases, support multi-hypothesis outputs, and provide better calibrated uncertainty, all of which can further strengthen reliability in sensitive scenarios.

\bibliography{custom}

@article{chu2024qwen2,
  title={Qwen2-audio technical report},
  author={Chu, Yunfei and Xu, Jin and Yang, Qian and Wei, Haojie and Wei, Xipin and Guo, Zhifang and Leng, Yichong and Lv, Yuanjun and He, Jinzheng and Lin, Junyang and others},
  journal={arXiv preprint arXiv:2407.10759},
  year={2024}
}

@article{team2024gemini,
  title={Gemini 1.5: Unlocking multimodal understanding across millions of tokens of context},
  author={Google},
  journal={arXiv preprint arXiv:2403.05530},
  year={2024}
}

@INPROCEEDINGS{Elsayed2022,
  title={Speech emotion recognition using supervised deep recurrent system for mental health monitoring},
  author={Elsayed, Nelly and ElSayed, Zag and Asadizanjani, Navid and Ozer, Murat and Abdelgawad, Ahmed and Bayoumi, Magdy},  title={Beyond silent letters: Amplifying llms in emotion recognition with vocal nuances},
  author={Wu, Zehui and Gong, Ziwei and Ai, Lin and Shi, Pengyuan and Donbekci, Kaan and Hirschberg, Julia},
  journal={arXiv preprint arXiv:2407.21315},
  year={2024}
}

@article{Ramakrishnan2013,
  title={Speech emotion recognition approaches in human computer interaction},
  author={Ramakrishnan, Srinivasan and El Emary, Ibrahiem MM},
  journal={Telecommunication Systems},
  volume={52},
  pages={1467--1478},
  year={2013},
  publisher={Springer}
}

@misc{niu2024textemotionunveilingemotion,
      title={From Text to Emotion: Unveiling the Emotion Annotation Capabilities of LLMs}, 
      author={Minxue Niu and Mimansa Jaiswal and Emily Mower Provost},
      year={2024},
      eprint={2408.17026},
      archivePrefix={arXiv},
      primaryClass={cs.CL},
      url={https://arxiv.org/abs/2408.17026}, 
}

@inproceedings{feng2024foundation,
  title={Foundation model assisted automatic speech emotion recognition: Transcribing, annotating, and augmenting},
  author={Feng, Tiantian and Narayanan, Shrikanth},
  booktitle={ICASSP 2024-2024 IEEE International Conference on Acoustics, Speech and Signal Processing (ICASSP)},
  pages={12116--12120},
  year={2024},
  organization={IEEE}
}

@misc{hong2024aerllm,
  title={AER-LLM: Ambiguity-aware Emotion Recognition Leveraging Large Language Models},
  author={Hong, Xin and Gong, Yuan and Sethu, Vidhyasaharan and Dang, Ting},
  journal={arXiv preprint arXiv:2409.18339},
  year={2024}
}

@article{busso2008iemocap,
  title={IEMOCAP: Interactive emotional dyadic motion capture database},
  author={Busso, Carlos and Bulut, Murtaza and Lee, Chi-Chun and Kazemzadeh, Abe and Mower, Emily and Kim, Samuel and Chang, Jeannette N and Lee, Sungbok and Narayanan, Shrikanth S},
  journal={Language resources and evaluation},
  volume={42},
  pages={335--359},
  year={2008},
  publisher={Springer}
}

@inproceedings{dang2017investigation,
  title={An Investigation of Emotion Prediction Uncertainty Using Gaussian Mixture Regression.},
  author={Dang, Ting and Sethu, Vidhyasaharan and Epps, Julien and Ambikairajah, Eliathamby},
  booktitle={INTERSPEECH},
  pages={1248--1252},
  year={2017}
}

@inproceedings{Mower2009,
    author    = {Emily Mower and Athanasios Metallinou and Chul Min Lee and Ali Kazemzadeh and Carlos Busso and Sungbok Lee and Shrikanth Narayanan},
    title     = {Interpreting ambiguous emotional expressions},
    booktitle = {Proceedings of the International Conference on Affective Computing and Intelligent Interaction},
    year      = {2009},
    pages     = {1--8}
}

@article{snell2024scaling,
  title={Scaling llm test-time compute optimally can be more effective than scaling model parameters},
  author={Snell, Charlie and Lee, Jaehoon and Xu, Kelvin and Kumar, Aviral},
  journal={arXiv preprint arXiv:2408.03314},
  year={2024}
}

@article{bi2024forest,
  title={Forest-of-thought: Scaling test-time compute for enhancing LLM reasoning},
  author={Bi, Zhenni and Han, Kai and Liu, Chuanjian and Tang, Yehui and Wang, Yunhe},
  journal={arXiv preprint arXiv:2412.09078},
  year={2024}
}

@article{liu2025can,
  title={Can 1B LLM Surpass 405B LLM? Rethinking Compute-Optimal Test-Time Scaling},
  author={Liu, Runze and Gao, Junqi and Zhao, Jian and Zhang, Kaiyan and Li, Xiu and Qi, Biqing and Ouyang, Wanli and Zhou, Bowen},
  journal={arXiv preprint arXiv:2502.06703},
  year={2025}
}

@article{zhang2025and,
  title={What, how, where, and how well? a survey on test-time scaling in large language models},
  author={Zhang, Qiyuan and Lyu, Fuyuan and Sun, Zexu and Wang, Lei and Zhang, Weixu and Guo, Zhihan and Wang, Yufei and King, Irwin and Liu, Xue and Ma, Chen},
  journal={arXiv preprint arXiv:2503.24235},
  year={2025}
}

@article{muennighoff2025s1,
  title={s1: Simple test-time scaling},
  author={Muennighoff, Niklas and Yang, Zitong and Shi, Weijia and Li, Xiang Lisa and Fei-Fei, Li and Hajishirzi, Hannaneh and Zettlemoyer, Luke and Liang, Percy and Cand{\`e}s, Emmanuel and Hashimoto, Tatsunori},
  journal={arXiv preprint arXiv:2501.19393},
  year={2025}
}

@inproceedings{yang2018predicting,
  title={Predicting Arousal and Valence from Waveforms and Spectrograms Using Deep Neural Networks.},
  author={Yang, Zixiaofan and Hirschberg, Julia},
  booktitle={Interspeech},
  pages={3092--3096},
  year={2018}
}

@article{hu2024exploring,
  title={Exploring Large-Scale Language Models to Evaluate EEG-Based Multimodal Data for Mental Health},
  author={Hu, Yongquan and Zhang, Shuning and Dang, Ting and Jia, Hong and Salim, Flora D and Hu, Wen and Quigley, Aaron J},
  journal={WellComp co-located with UbiComp 2024},
  year={2024}
}

@article{wei2022chain,
  title={Chain-of-thought prompting elicits reasoning in large language models},
  author={Wei, Jason and Wang, Xuezhi and Schuurmans, Dale and Bosma, Maarten and Xia, Fei and Chi, Ed and Le, Quoc V and Zhou, Denny and others},
  journal={Advances in neural information processing systems},
  volume={35},
  pages={24824--24837},
  year={2022}
}

@article{fu2022complexity,
  title={Complexity-based prompting for multi-step reasoning},
  author={Fu, Yao and Peng, Hao and Sabharwal, Ashish and Clark, Peter and Khot, Tushar},
  journal={arXiv preprint arXiv:2210.00720},
  year={2022}
}

@article{shu2023audio,
  title={Audio-visual llm for video understanding},
  author={Shu, Fangxun and Zhang, Lei and Jiang, Hao and Xie, Cihang},
  journal={arXiv preprint arXiv:2312.06720},
  year={2023}
}

@article{wang2022self,
  title={Self-consistency improves chain of thought reasoning in language models},
  author={Wang, Xuezhi and Wei, Jason and Schuurmans, Dale and Le, Quoc and Chi, Ed and Narang, Sharan and Chowdhery, Aakanksha and Zhou, Denny},
  journal={arXiv preprint arXiv:2203.11171},
  year={2022}
}

@article{lotfian2017building,
  title={Building naturalistic emotionally balanced speech corpus by retrieving emotional speech from existing podcast recordings},
  author={Lotfian, Reza and Busso, Carlos},
  journal={IEEE Transactions on Affective Computing},
  volume={10},
  number={4},
  pages={471--483},
  year={2017},
  publisher={IEEE}
}

@article{cao2014crema,
  title={Crema-d: Crowd-sourced emotional multimodal actors dataset},
  author={Cao, Houwei and Cooper, David G and Keutmann, Michael K and Gur, Ruben C and Nenkova, Ani and Verma, Ragini},
  journal={IEEE transactions on affective computing},
  volume={5},
  number={4},
  pages={377--390},
  year={2014},
  publisher={IEEE}
}

@article{xu2025qwen2,
  title={Qwen2. 5-omni technical report},
  author={Xu, Jin and Guo, Zhifang and He, Jinzheng and Hu, Hangrui and He, Ting and Bai, Shuai and Chen, Keqin and Wang, Jialin and Fan, Yang and Dang, Kai and others},
  journal={arXiv preprint arXiv:2503.20215},
  year={2025}
}

@article{Qwen2-Audio,
  title={Qwen2-Audio Technical Report},
  author={Chu, Yunfei and Xu, Jin and Yang, Qian and Wei, Haojie and Wei, Xipin and Guo,  Zhifang and Leng, Yichong and Lv, Yuanjun and He, Jinzheng and Lin, Junyang and Zhou, Chang and Zhou, Jingren},
  journal={arXiv preprint arXiv:2407.10759},
  year={2024}
}

@article{ghosh2025audio,
  title={Audio Flamingo 2: An audio-language model with long-audio understanding and expert reasoning abilities},
  author={Ghosh, Sreyan and Kong, Zhifeng and Kumar, Sonal and Sakshi, S and Kim, Jaehyeon and Ping, Wei and Valle, Rafael and Manocha, Dinesh and Catanzaro, Bryan},
  journal={arXiv preprint arXiv:2503.03983},
  year={2025}
}

@article{achiam2023gpt,
  title={Gpt-4 technical report},
  author={Achiam, Josh and Adler, Steven and Agarwal, Sandhini and Ahmad, Lama and Akkaya, Ilge and Aleman, Florencia Leoni and Almeida, Diogo and Altenschmidt, Janko and Altman, Sam and Anadkat, Shyamal and others},
  journal={arXiv preprint arXiv:2303.08774},
  year={2023}
}

@article{team2023gemini,
  title={Gemini: a family of highly capable multimodal models},
  author={Team, Gemini and Anil, Rohan and Borgeaud, Sebastian and Alayrac, Jean-Baptiste and Yu, Jiahui and Soricut, Radu and Schalkwyk, Johan and Dai, Andrew M and Hauth, Anja and Millican, Katie and others},
  journal={arXiv preprint arXiv:2312.11805},
  year={2023}
}

@article{sethu2019ambiguous,
  title={The ambiguous world of emotion representation},
  author={Sethu, Vidhyasaharan and Provost, Emily Mower and Epps, Julien and Busso, Carlos and Cummins, Nicholas and Narayanan, Shrikanth},
  journal={arXiv preprint arXiv:1909.00360},
  year={2019}
}

@inproceedings{dudzik2024indeterminacy,
  title={Indeterminacy in Affective Computing: Considering Meaning and Context in Data Collection Practices},
  author={Dudzik, Bernd and Hrkalovic, Tiffany Matej and Hao, Chenxu and Raman, Chirag and Tsfasman, Masha},
  booktitle={2024 12th International Conference on Affective Computing and Intelligent Interaction Workshops and Demos (ACIIW)},
  pages={181--185},
  year={2024},
  organization={IEEE}
}

@article{adler2024nemotron,
  title={Nemotron-4 340b technical report},
  author={Adler, Bo and Agarwal, Niket and Aithal, Ashwath and Anh, Dong H and Bhattacharya, Pallab and Brundyn, Annika and Casper, Jared and Catanzaro, Bryan and Clay, Sharon and Cohen, Jonathan and others},
  journal={arXiv preprint arXiv:2406.11704},
  year={2024}
}

@inproceedings{wu2024emotion,
  title={Emotion Recognition Systems Must Embrace Ambiguity},
  author={Wu, Jingyao and Dang, Ting and Sethu, Vidhyasaharan and Ambikairajah, Eliathamby},
  booktitle={2024 12th International Conference on Affective Computing and Intelligent Interaction Workshops and Demos (ACIIW)},
  pages={166--170},
  year={2024},
  organization={IEEE}
}

@inproceedings{wu24_interspeech,
  title     = {Dual-Constrained Dynamical Neural ODEs for Ambiguity-aware Continuous Emotion Prediction},
  author    = {Jingyao Wu and Ting Dang and Vidhyasaharan Sethu and Eliathamby Ambikairajah},
  year      = {2024},
  booktitle = {Interspeech 2024},
  pages     = {3185--3189},
  doi       = {10.21437/Interspeech.2024-119},
  issn      = {2958-1796},
}

@misc{ultravox_v03,
  author       = {{Fixie AI}},
  title        = {A fast multimodal LLM designed for real-time voice interactions},
  year         = {2024},
  howpublished = {\url{https://github.com/fixie-ai/ultravox/releases/tag/v0.3}},
  note         = {Accessed: 2025-07-31}
}

@misc{ultravox_v04,
  author       = {{Fixie AI}},
  title        = {A fast multimodal LLM designed for real-time voice interactions},
  year         = {2024},
  howpublished = {\url{https://github.com/fixie-ai/ultravox/releases/tag/v0.4}},
  note         = {Accessed: 2025-07-31}
}

@techreport{gemini2flash,
  author       = {{Google}},
  title        = {Gemini 2.0 Flash - Model Card},
  institution  = {Google},
  year         = {2025},
  month        = {April},
  url          = {https://storage.googleapis.com/model-cards/documents/gemini-2-flash.pdf}
}

@article{abd2019overview,
  title={An overview of the features of chatbots in mental health: A scoping review},
  author={Abd-Alrazaq, Alaa A and Alajlani, Mohannad and Alalwan, Ali Abdallah and Bewick, Bridgette M and Gardner, Peter and Househ, Mowafa},
  journal={International journal of medical informatics},
  volume={132},
  pages={103978},
  year={2019},
  publisher={Elsevier}
}

@article{vistorte2024integrating,
  title={Integrating artificial intelligence to assess emotions in learning environments: a systematic literature review},
  author={Vistorte, Angel Olider Rojas and Deroncele-Acosta, Angel and Ayala, Juan Luis Mart{\'\i}n and Barrasa, Angel and L{\'o}pez-Granero, Caridad and Mart{\'\i}-Gonz{\'a}lez, Mariacarla},
  journal={Frontiers in psychology},
  volume={15},
  pages={1387089},
  year={2024},
  publisher={Frontiers Media SA}
}

@article{flynn2020assessing,
  title={Assessing the effectiveness of automated emotion recognition in adults and children for clinical investigation},
  author={Flynn, Maria and Effraimidis, Dimitris and Angelopoulou, Anastassia and Kapetanios, Epaminondas and Williams, David and Hemanth, Jude and Towell, Tony},
  journal={Frontiers in human neuroscience},
  volume={14},
  pages={70},
  year={2020},
  publisher={Frontiers Media SA}
}

@inproceedings{wu2022novel,
  title={A novel sequential monte carlo framework for predicting ambiguous emotion states},
  author={Wu, Jingyao and Dang, Ting and Sethu, Vidhyasaharan and Ambikairajah, Eliathamby},
  booktitle={ICASSP 2022-2022 IEEE International Conference on Acoustics, Speech and Signal Processing (ICASSP)},
  pages={8567--8571},
  year={2022},
  organization={IEEE}
}

@inproceedings{han2017hard,
  title={From hard to soft: Towards more human-like emotion recognition by modelling the perception uncertainty},
  author={Han, Jing and Zhang, Zixing and Schmitt, Maximilian and Pantic, Maja and Schuller, Bj{\"o}rn},
  booktitle={Proceedings of the 25th ACM international conference on Multimedia},
  pages={890--897},
  year={2017}
}

@article{wu2024handling,
  title={Handling ambiguity in emotion: From out-of-domain detection to distribution estimation},
  author={Wu, Wen and Li, Bo and Zhang, Chao and Chiu, Chung-Cheng and Li, Qiujia and Bai, Junwen and Sainath, Tara N and Woodland, Philip C},
  journal={arXiv preprint arXiv:2402.12862},
  year={2024}
}

@article{halim2025token,
  title={Token-Level Logits Matter: A Closer Look at Speech Foundation Models for Ambiguous Emotion Recognition},
  author={Halim, Jule Valendo and Wang, Siyi and Jia, Hong and Dang, Ting},
  journal={INTERSPEECH},
  year={2025}
}

@article{liu2025rethinking,
  title={Rethinking the Role of Prompting Strategies in LLM Test-Time Scaling: A Perspective of Probability Theory},
  author={Liu, Yexiang and Li, Zekun and Fang, Zhi and Xu, Nan and He, Ran and Tan, Tieniu},
  journal={arXiv preprint arXiv:2505.10981},
  year={2025}
}

\appendix

\appendix
\setcounter{figure}{0} 
\setcounter{table}{0} 

\clearpage
\section{Technical Appendices and Supplementary Material}
\subsection{Prompting}\label{app:prompt}
The example prompts used for audio language models (ALMs) without CoT are shown in Table~\ref{tab:llmp}. The Background section provides the model with the fundamental scenario. The Task section explicitly instructs the model to predict the probability distribution of emotions for the target utterance from a predefined list of categories. Finally, the Output Constraints section restricts the model's output to a specific format to facilitate subsequent automated evaluation.
\par
In the context of the CREMA-D dataset, the emotion is conveyed through the actor's delivery rather than the semantic content of the 12 fixed target utterances. Therefore, to provide a clearer directive to ALMs and improve results, we revised the prompt. Specifically, we changed it from “Predict the probability distribution of emotions for the target utterance” to “Predict the probability distribution of emotions for the audio".

\begin{table}[ht!]
    \centering
    \small
    \caption{Prompt template for discrete emotion classification on the IEMOCAP dataset.}
    \label{tab:llmp}
    \renewcommand{\arraystretch}{1.3} 
    \resizebox{\columnwidth}{!}{%
    \begin{tabular}{p{1.5cm}|p{6.5cm}}
        \toprule
        \textbf{Background} & 
        Two speakers are having a conversation. \\
        \hline
        \textbf{Target \newline Utterance} & 
        \texttt{Yeah. I suppose I have been. But it's going from me.} \\
        \hline
        \textbf{Task} & 
        Predict the probability distribution of emotions for the target utterance from the following options: \texttt{['Neutral state', 'Happiness', 'Anger', 'Sadness']}.
        \newline\newline
        Consider both the context and acoustic features. \\
        \hline
        \textbf{Output \newline Constraints} & 
        \begin{enumerate}
            \item Generate EXACTLY this JSON structure: \newline \texttt{\{"Neutral state":float, "Happiness":float, "Anger":float, "Sadness":float\}} \newline \textit{Before outputting, check if the format of your output is in accordance with the requirements I provided.}
            \item Sum of probabilities must equal to 1.0.
            \item Do not include any explanations or text besides the dictionary.
        \end{enumerate} \\
        \bottomrule
    \end{tabular}
    }
\end{table}

The CoT prompting additionally introduced the step-by-step instructions as shown in Table~\ref{tab:cot_prompt}. The prompt design for ALM-Verifier is shown in Table~\ref{tab:alm-prompt}.
\begin{table}[ht!]
    \centering
    \small
    \caption{Emotion recognition prompt template with CoT}
    \label{tab:cot_prompt}
    \renewcommand{\arraystretch}{1.3} 
    \resizebox{\columnwidth}{!}{%
    \begin{tabular}{p{1.5cm}|p{6.5cm}}
        \toprule
        
        \textbf{Background} & 
        Two speakers are having a conversation. \\
        \hline
        
        \textbf{Target \newline Utterance} & 
        \texttt{Yeah. I suppose I have been. But it's going from me.} \\
        \hline
        
        \textbf{Task} & 
        Predict the probability distribution of emotions for the target utterance from the following options: \texttt{['Neutral state', 'Happiness', 'Anger', 'Sadness']}.
        \newline\newline
        Consider both the context and acoustic features. \\
        \hline

        \textbf{CoT \newline Instructions} & 
        \begin{itemize}
            \item \textbf{First}, carefully listen to the tone, intonation, pauses, vocal energy, and other acoustic features.
            \item \textbf{Second}, examine the textual content, considering words, sentiment, and contextual cues.
            \item Identify all emotional cues, even subtle ones.
            \item For each emotion, evaluate if it is present. If multiple are present, estimate their relative strength and presence.
        \end{itemize} \\
        \hline
        
        \textbf{Output \newline Constraints} & 
        \begin{enumerate}
            \item Generate EXACTLY this JSON structure: \newline \texttt{\{"Neutral state":float, ...\}} \newline \textit{Before outputting, check if the format of your output is in accordance with the requirements I provided.}
            \item Sum of probabilities must equal to 1.0.
            \item Do not include any explanations or text besides the dictionary.
        \end{enumerate} \\
        \bottomrule
    \end{tabular}
    }
\end{table}

\begin{table}[ht!]
    \centering
    \small
    \caption{Emotion recognition prompt template with ALM-Verifier}
    \label{tab:alm-prompt}
    \renewcommand{\arraystretch}{1.3} 
    \resizebox{\columnwidth}{!}{%
    \begin{tabular}{p{1.5cm}|p{6.5cm}}
        \toprule
        
        \textbf{Background} & 
        You are evaluating a model’s prediction for an emotion distribution task. \\
        \hline
        \textbf{Role} & 
        User \\
        \hline
        \textbf{User Prompt} & 
        Predict the probability distribution of emotions for the target utterance from the following options: \texttt{['Neutral state', 'Happiness', 'Anger', 'Sadness']}.
        \newline\newline
        Consider both the context and acoustic features. \\
        \hline

    \textbf{Model Output} & 
        \texttt{\{"Anger": 0.1, "Happy": 0.5, ...\}} \\
        \hline
        
        \textbf{Task} & 
       Only scoring based on emotion content: how well does the distribution reflect the emotional content of the utterance and context?\\
       \hline
        \textbf{Scoring \newline Guide} & 
        \begin{itemize}
            \item \textbf{0.0–0.3}: Clearly wrong or unrelated

            \item \textbf{0.4–0.7}: Partially reasonable or somewhat mismatched
            \item \textbf{0.8–1.0}: Good match to the expected emotional distribution
            \item Reply ONLY with a number between 0.0 and 1.0.
        \end{itemize} \\

        \bottomrule
    \end{tabular}
    }
\end{table}

\subsection{Implementation details}

\paragraph{Model structures}\label{app:model}
The models used in this study includes seven audio LLMs.
\begin{itemize}
\item \textbf{Qwen2.5-Omni}~\cite{xu2025qwen2}: an advanced open-source multimodal model by Alibaba, is designed to seamlessly process both audio, visual and text input. It supports audio understanding, speech interaction, and cross-modal reasoning. In this study, we utilize the 7B parameter variant, which demonstrates strong performance in various audio-language integration tasks.

\item \textbf{Qwen2-Audio-Instruct}~\citep{chu2024qwen2}: Created by Alibaba, Qwen2-Audio is a cutting-edge audio-text model designed to interpret and respond to spoken instructions. It processes diverse audio inputs for analysis or text generation. Our experiments employed the 7-billion-parameter version of this model.
\par

\item \textbf{Audio-Flamingo 2}~\citep{ghosh2025audio}: NVIDIA’s Audio-Flamingo 2 builds on the original Flamingo architecture with improved capacity for reasoning over extended audio segments. Despite utilizing a relatively compact 3B parameter language core, it surpasses many peers across more than 20 standard benchmarks in audio understanding.

\item \textbf{Ultravox-v0.3}~\citep{ultravox_v03}: Developed by Fixie AI, Ultravox-v0.3 is a large-scale speech-to-text model engineered for low-latency, real-time transcription and speech understanding. Built upon a modified Llama-3 8B architecture, it excels at processing long-form audio and accurately transcribing speech even in noisy conditions, making it suitable for advanced conversational AI applications.

\item \textbf{Ultravox-v0.4}~\citep{ultravox_v04}: As an advancement over its predecessor, Fixie AI's Ultravox-v0.4 further enhances transcription accuracy and robustness. This version introduces architectural refinements that significantly reduce the Word Error Rate (WER) and improve handling of diverse accents and acoustic environments. It retains the core strengths of real-time processing while offering superior performance for complex speech understanding tasks.

\item \textbf{Gemini-2.0-Flash}~\citep{gemini2flash}: As a streamlined variant of the Gemini-2.0 family introduced by Google in 2025, Gemini-2.0-Flash is tailored for low-latency audio processing. It offers efficient speech-related functionalities such as recognition and synthesis while operating with reduced computational overhead.

\item \textbf{Gemini-2.5-Pro}~\citep{team2024gemini}: This model represents a high-capacity iteration within the Gemini-1.5 lineup, designed for multimodal tasks involving audio. With expanded training data and architectural refinements, it achieves strong results in domains like emotion detection, multilingual audio interpretation, and speech-to-text transformation.

\item \textbf{GPT-4o}~\citep{achiam2023gpt}: OpenAI’s GPT-4o introduces a unified framework for handling both audio and text inputs. Its capabilities span a range of tasks—from natural-sounding speech synthesis to real-time transcription and conversational reasoning—making it a versatile model for cross-modal interaction.

\end{itemize}

\paragraph{Output processing}
Notably, the outputs from Audio-Flamingo 2 did not consistently adhere to our predefined format, a limitation we attribute to its smaller model size. Consequently, we implemented some strategies shown in Table~\ref{tab:parse_model_output}, which resulted in only 57.9\%, 66.5\%, and 87.2\% of the data being valid for the IEMOCAP, MSP-Podcast, and CREMA-D datasets, respectively. In contrast, all other models produced fully compliant outputs, allowing for the use of their complete datasets.

\begin{table}[t!]
    \centering
    \small
    \caption{Parsing Strategies }
    \label{tab:parse_model_output}
    \renewcommand{\arraystretch}{1.3}
    \resizebox{\columnwidth}{!}{%
    \begin{tabular}{p{1.1cm}|p{6.6cm}}
        \hline
        \textbf{Strategy} & \textbf{Description} \\
        \hline

        \textbf{JSON Dictionary} &
        Attempts to parse the string as a JSON object like \texttt{\{"Anger": 0.1, "Happy": 0.5, ...\}}. If successful, it converts the dictionary into a probability vector and normalizes it. \\
        \hline

        \textbf{Emotion List} &
        Handles raw strings like \texttt{['disgust', 'neutral']} or \texttt{["happy"]}. The matched emotions are mapped to their indices, and a uniform distribution is created over them. Unrecognized entries are ignored. \\
        \hline

        \textbf{Float List} &
        Detects numeric sequences inside square brackets (e.g., \texttt{[0.3, 0.2, 0.5]}). If at least two values are found, it is interpreted as a probability vector. The vector is then padded or trimmed to fit the expected length and normalized. \\
        \hline

        \textbf{Keyword Match} &
        Searches for direct mentions of known emotion keywords (e.g., \texttt{sad}, \texttt{anger}) in the raw string. If found, a one-hot vector is returned with the matched emotion set to 1.0. \\
        \hline

        \textbf{Single Float} &
        If the raw output contains only a single number or a single-element list (e.g., \texttt{0.54} or \texttt{[0.54]}), it is not treated as a valid emotion distribution and is discarded. \\
        \hline
    \end{tabular}
    }
\end{table}

\begin{figure*}[t]
  \centering
  \includegraphics[width=\textwidth]{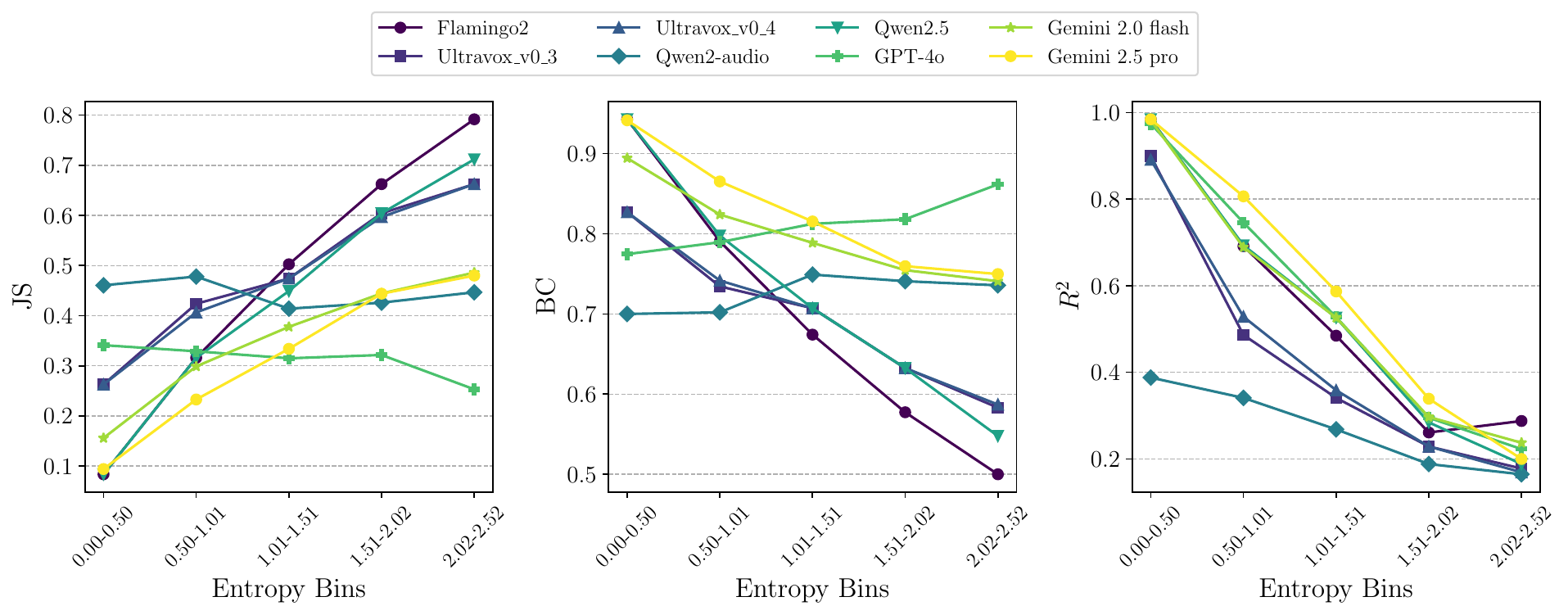}
  \caption{Performance of ALMs on CREMA-D dataset across different entropy bins. This analysis presents the median to mitigate the influence of occasional outliers on the overall model performance evaluation.}
  \label{fig:ALM_line_plot}
\end{figure*}

\subsection{Further analysis}
\paragraph{Performance of ALMs with varying difficulty levels}

The performance hierarchy of the models generally aligns with the analysis in Table \ref{tab:alms}. Closed-source models, as a group, outperform their open-source counterparts, with Gemini 2.5 Pro emerging as the top-performing model overall. Of particular interest is the remarkable robustness demonstrated by Qwen2-Audio and GPT-4o on the JS Divergence and Bhattacharyya Coefficient metrics. While the performance of other models degraded with increasing data entropy, Qwen2-Audio's scores remained nearly constant, and GPT-4o's even showed a counter-intuitive improvement. This suggests a superior capability of these two models to interpret complex emotional states.

\end{document}